# Pulse Shape Analysis with a Broad-Energy Germanium Detector for the GERDA experiment

Dušan Budjáš, Marik Barnabé Heider, Oleg Chkvorets, Stefan Schönert, Nikita Khanbekov

*Abstract–* To reduce background in experiments looking for rare events, such as the GERDA double beta decay experiment, it is necessary to employ active background-suppression techniques. One of such techniques is the pulse shape analysis of signals induced by the interaction of radiation with the detector. Analysis of the time-development of the impulses can distinguish between an interaction of an electron and an interaction of a multiple-scattered photon inside the detector. This information can be used to eliminate background events from the recorded data. Results of pulse-shape analysis of signals from a commercially available broad-energy germanium detector are presented and the pulse-shape discrimination capability of such detector configuration for use in low-background experiments is discussed.

## I. Introduction

The GERDA (GERmanium Detector Array) experiment [1] aims to search for the neutrinoless double-beta ($0\nu\beta\beta$) decay of $^{76}$Ge. The main design feature of GERDA is the use of liquid argon as a shield against gamma radiation [2], the dominant background in earlier experiments [3, 4]. High purity germanium detectors are immersed directly into the cryogenic liquid which also acts as a cooling medium.

In the first phase of the experiment, 18 kg of Ge-detectors enriched in $^{76}$Ge, which were previously operated by the Heidelberg-Moscow and IGEX collaborations, will be redeployed. The background level in the germanium diodes is envisioned to be $< 10^{-2}$ counts/(keV·kg·y) in the region of interest ($Q_{\beta\beta}$ = 2039 keV), more than one order of magnitude lower than in previous $0\nu\beta\beta$-decay experiments. The second phase, for which new enriched Ge-detectors (additional 20 kg) will be custom made, aims at another order of magnitude reduction in background to a level of $< 10^{-3}$ cts/(keV·kg·y).

To reach such low background levels, active background-suppression methods are necessary in addition to the low-background design of the experiment. Among them, an analysis of the time-structure of the Ge-detector response will be applied. Pulse shape analysis methods have already been used in previous double-beta decay experiments [5-9]. The goal of these techniques is to distinguish between $\beta\beta$-decay events, which are created by localised interactions of the emitted electrons, and background events, which are predominantly photon-induced. Interaction of a photon inside a Ge-crystal is likely to happen via multiple scatterings. Subsequent charge collection on the signal electrode results in different time-structure of the signal than in the case of electron-like single-site interaction.

The detector configuration under study (Fig. 1) has potentially superior ability to distinguish such single- and multi-site interactions, compared to usual coaxial detectors. The signal is read-out from a small area $p^+$ contact. Charge collection on such a small electrode and the specific electric field distribution in this type of detector enhance the differences in charge-carrier drift times depending on the site of interaction, and thus offer better discrimination power of pulse-shape analysis [10]. An additional advantage pointed out in previous investigations of similar detector configurations [10, 11] is its low capacitance, resulting in lower noise and therefore improved resolution at low energies.

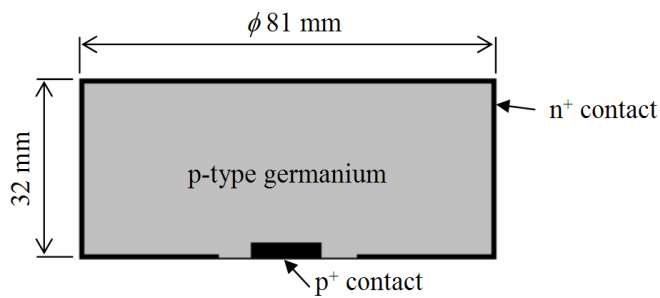

Fig. 1. Schematic drawing of the studied 878 g p-type crystal of a broad-energy Ge-detector, manufactured by Canberra Semiconductor, Olen [12].

## II. Experimental Setup

The broad-energy Ge-detector (BEGe) was purchased from Canberra at the beginning of 2008. The front-end electronics and data acquisition system (DAQ) consisted of Canberra 2002CSL preamplifier, with specified noise of 570 eV (at 0 pF input capacitance) and rise time of 20 ns; Canberra model 2111 analogue timing-filter amplifier (TFA); a non-shaping amplifier built in-house; and Struck SIS 3301 14-bit, 100 MHz flash-ADC. The flash-ADC was used for producing energy spectrum via digital shaping (10 μs shaping constant), using amplified "energy" output of the preamplifier, and for recording pulse shapes after 10 ns differentiation and 10 ns integration by the TFA (using the "timing" output). Examples of recorded signals are on Fig. 2.

Manuscript received December 9, 2008. This work was supported by the Transregio Sonderforschungsbereich SFB/TR27 'Neutrinos and Beyond' by the Deutsche Forschungsgemeinschaft and by the INTAS grant 1000008-7996 of the European Commission.

Authors are with the Max-Planck-Institut für Kernphysik, Saupfercheckweg 1, D-69117 Heidelberg, Germany (corresponding author: Dušan Budjáš, Tel. +49 6221 516 827, Fax +49 6221 516 872, E-mail: dusan.budjas@mpi-hd.mpg.de).

O. Chkvorets is now with the Department of Physics, Laurentian University, Ramsey Lake Road, P3E 2C6 Sudbury, Ontario, Canada.

N. Khanbekov is also with the Institute for Theoretical and Experimental Physics, Bolshaya Cheremushkinskaya 25, 117218 Moscow, Russia.

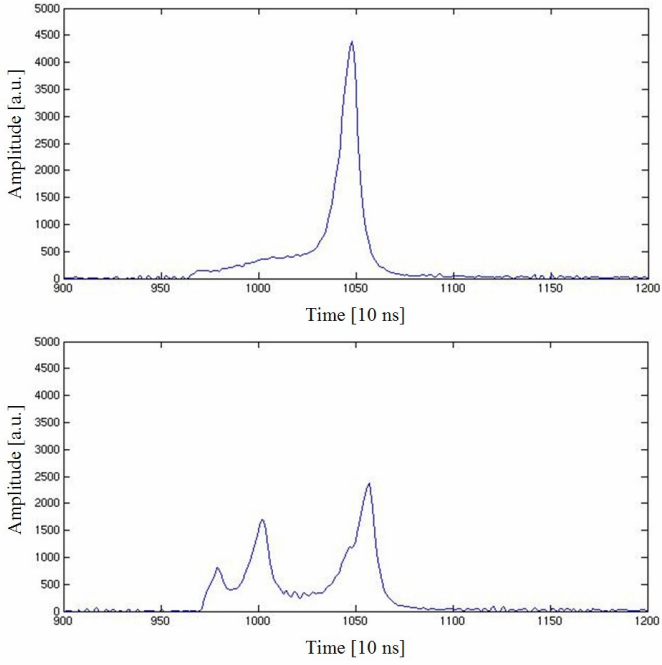

Fig. 2. Comparison of signal traces of a typical electron event (top) and a typical multiple-scattered γ-ray interaction (bottom) after differentiation by TFA. The energy of both events was equal.

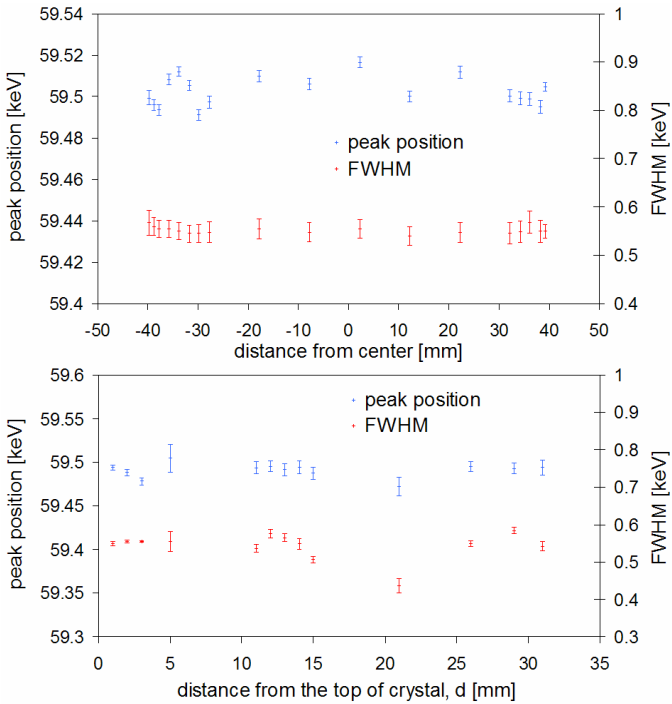

Fig. 3. Peak position and energy resolution variation along the top (upper plot) and side (lower plot) surfaces of the BEGe detector, obtained with a collimated 59.5 keV γ-ray beam from $^{241}$Am source. The top scan shows maximal peak position variation of 0.075% and the side scan of 0.055%. Such gain variations are consistent with the noise-induced gain fluctuations observed in the employed DAQ system.

The energy resolution of the system was 0.5 keV FWHM at 59.5 keV and 1.6 keV FWHM at 1332.5 keV. Detector surface scanning with collimated 59.5 keV γ-ray beam was performed to verify homogeneity and isotropy of the detector response (Fig. 3.). The change of the 59.5 keV peak position in the measured spectra was used as an indicator of gain variation along the surface of the detector. Maximal variation was 0.075%, in agreement with the observed gain fluctuation caused by the noise of the DAQ system. The dead layer thickness at the top surface of the crystal was measured at $(0.42 \pm 0.01)$ mm. The active mass of the detector was determined to be $(836 \pm 9)$ g, 95% of total. The dead layer thickness and the active mass were evaluated using a $^{241}$Am, respectively a $^{60}$Co source, according to the procedure described in [13]. No indication of incomplete charge collection was observed in these tests.

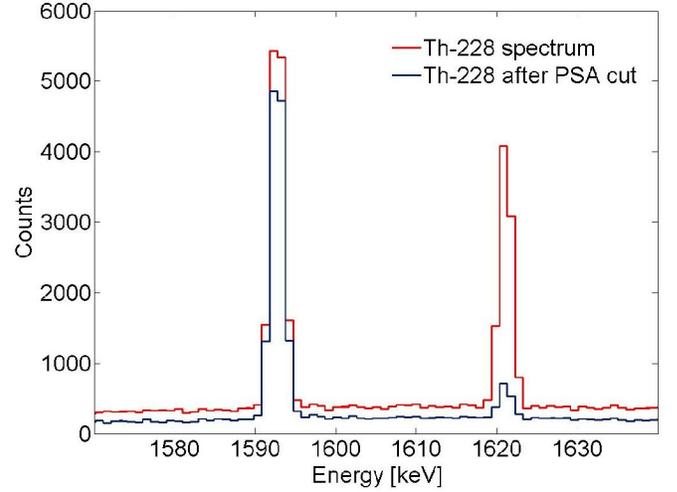

Fig. 4. Spectrum of $^{228}$Th before and after single-site/multi-site pulse-shape discrimination. The 1592.5 keV double-escape peak from the 2614.5 keV emission line of $^{208}$Tl contains mostly electron and positron absorption events, approximately analogous to ββ-decay events. Differences are in the energy of the DEP electrons/positrons (78% of the energy of ββ-decay electrons) and different spatial distributions of interactions inside the crystal. The resulting changes to the ββ-decay signal acceptance are expected to be small. The 1620.5 keV line is a full-energy γ-ray peak, containing mostly multiple-scattering interactions of photons with subsequent absorption.

III. RESULTS

Single-parameter discrimination, based on the normalised amplitude of the signal trace passed through the TFA, was performed to distinguish electron and multiple-scattered photon signals. The 1592.5 keV double-escape peak from the 2614.5 keV emission line of $^{208}$Tl was used as a sample of electron-induced events, while full-energy γ-ray peaks were used as samples containing mostly multiple-scattered events. The fractions of events remaining after the single-parameter cut in the peaks (after subtracting Compton-continuum background) are shown in Table I. The 2103.5 keV line is the single-escape peak from the 2614.5 keV γ-rays and contains dominantly multiple-site interactions (the localised $e^{-}$ and $e^{+}$ absorption after the pair-production event, followed by the scatterings and absorption of one of the annihilation γ-rays). The lower part of the table contains the fractions remaining in the region around the Q-value of the $^{76}$Ge ββ-decay in spectra from $^{228}$Th, $^{226}$Ra and $^{60}$Co. Example part of the $^{228}$Th spectrum before and after the discrimination is shown on Fig. 4.

While the events in the double-escape peak of $^{208}$Tl do not simulate exactly the expected signal from $0\nu\beta\beta$-decay, the difference in accepted fraction of events is expected to be relatively small. The isotopes measured represent the kind of backgrounds expected in the materials surrounding the detectors in the GERDA and similar ultra-low background experiments.

TABLE I
FRACTIONS OF EVENTS REMAINING AFTER SINGLE-PARAMETER DISCRIMINATION OF RECORDED SIGNAL TRACES

| Region | Remaining fraction |
|---|---|
| 1592.5 ± 6.1 keV | 90.9% ± 1.4% |
| 1620.5 ± 6.1 keV | 12.5% ± 0.8% |
| 2103.5 ± 10.2 keV | 8.5% ± 0.5% |
| 2614.5 ± 7.7 keV | 13.2% ± 0.1% |
| 2039 ± 14 keV ($^{60}$Co) | 1.69% ± 0.06% |
| 2039 ± 14 keV ($^{226}$Ra) | 31.8% ± 0.9% |
| 2039 ± 14 keV ($^{228}$Th) | 49.2% ± 0.8% |

IV. CONCLUSIONS

The results of the first pulse-shape discrimination of electron and multiple-scattered photon events with BEGe detector indicate very good background reduction power for application in the GERDA experiment, while keeping large acceptance of the $0\nu\beta\beta$-decay signal. The performance is similar to that of highly segmented Ge-detectors [14, 15]. The combination of its background discrimination power, excellent energy resolution and low demands on the amount of signal readouts (thus reducing potential background sources around detectors and the complexity of DAQ systems) make the BEGe type of detector well suitable for many ultra-low background applications. In addition, the established commercial practice of BEGe crystal production at O(1 kg) masses offers the prospect of reduced amount of R&D efforts needed to bring this type of detectors into the scientific use in physics experiments.

ACKNOWLEDGMENT

We thank Dr. Jan Verplancke, Canberra Semiconductor N.V. Olen, who pointed out to us that BEGe detectors might have similar field configurations and thus pulse shape properties as those described in ref. [10].